%% file: main.tex
\documentclass[journal]{vgtc}                

\ifpdf
  \pdfoutput=1\relax                   
  \usepackage{graphicx}                
  \DeclareGraphicsExtensions{.pdf,.png,.jpg,.jpeg} 
\else
  \usepackage{graphicx}                
  \DeclareGraphicsExtensions{.eps}     
\fi%

\graphicspath{{figures/}{pictures/}{images/}{./}} 

\usepackage{microtype}                 
\PassOptionsToPackage{warn}{textcomp}  
\usepackage{textcomp}                  
\usepackage{mathptmx}                  
\usepackage{times}                     
\usepackage{cite}                      
\usepackage{tabu}                      
\usepackage{booktabs}                  

\usepackage{xcolor}
\usepackage{balance}
\usepackage{float}
\usepackage{enumitem}
\usepackage[indentfirst=false,rightmargin=5pt]{quoting}
\quotingsetup{vskip=2pt}

\definecolor{orange}{rgb}{1, 0.55, 0.0}

\newcommand{\revis}[1]{{\textcolor{black}{#1}}}
\usepackage{url}

\onlineid{0}

\vgtccategory{Research}
\vgtcpapertype{evaluation}

\title{Learning Objectives, Insights, and Assessments: \\ How Specification Formats Impact Design}

\author{Elsie Lee-Robbins, Shiqing He, and Eytan Adar}
\authorfooter{
\item
 Elsie Lee-Robbins, Shiqing He, and Eytan Adar are with University of Michigan, School of Information. E-mail: elsielee,heslicia,eadar@umich.edu.
 }

\shortauthortitle{Lee-Robbins \MakeLowercase{\textit{et al.}}: Insights, Learning Objectives, and Assessments: How Specification Formats Impact Design}

\abstract{
    \input{sections/00_abstract}
}
   

\CCScatlist{ 
 \CCScat{K.6.1}{Management of Computing and Information Systems}%
{Project and People Management}{Life Cycle};
 \CCScat{K.7.m}{The Computing Profession}{Miscellaneous}{Ethics}
}

\teaser{
  \centering
  \includegraphics[width=\linewidth]{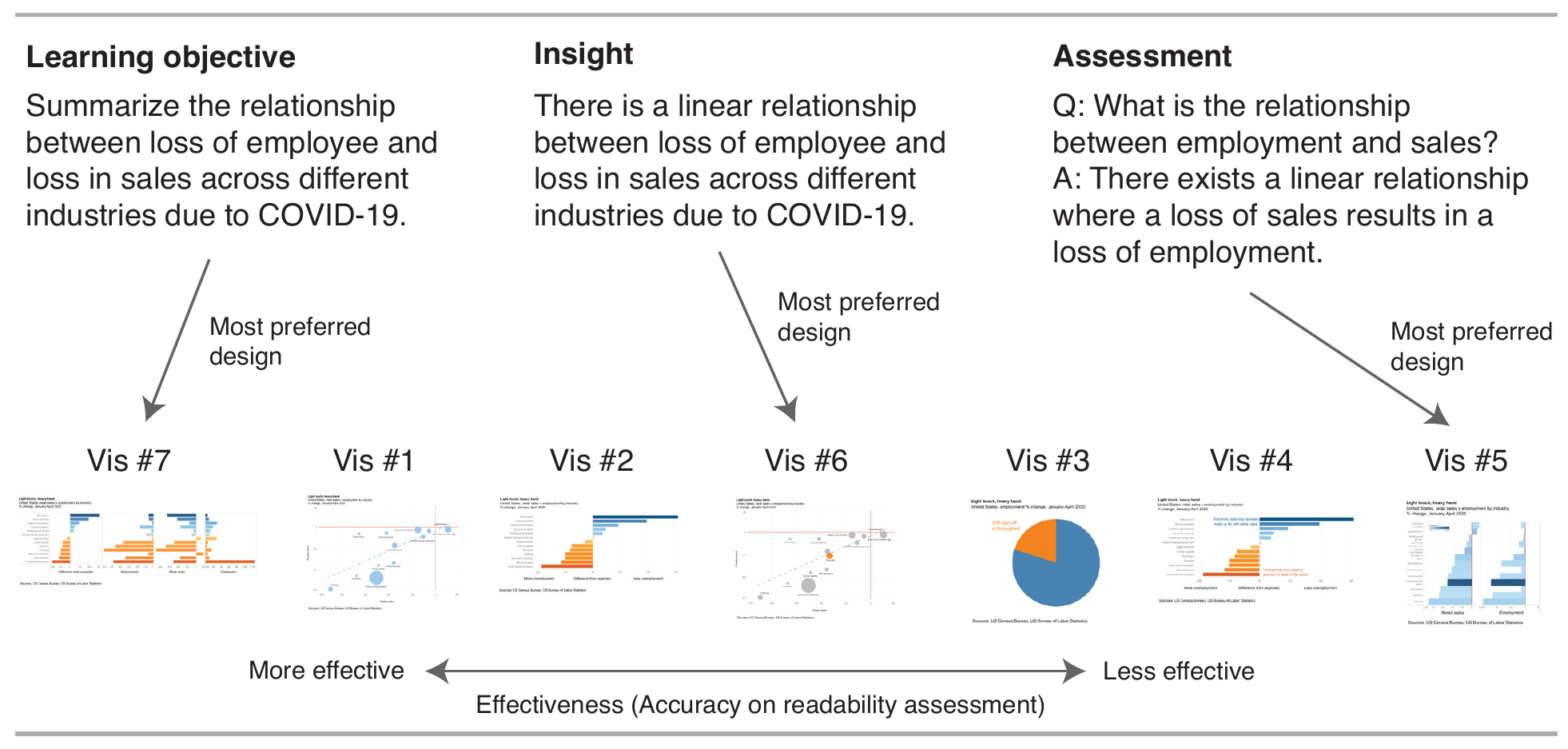}
    \caption{Example of three specifications (learning objective, insight, and assessment) with the most preferred visualization design identified for each specification. 
    The seven visualization designs are ordered from most effective to least effective.}
	\label{fig:teaser}
}

\vgtcinsertpkg

\begin{document}



\maketitle

\input{sections/01_introduction}
\input{sections/02_related}
\input{sections/03_intent}
\input{sections/04_study}

\input{sections/05_result}

\input{sections/06_discussion}

\input{sections/07_conclusion}
\input{sections/08_ack}
\balance

\bibliographystyle{abbrv-doi.bst}

\bibliography{bib.bib}

\end{document}

%% file: sections/00_abstract.tex
Despite the ubiquity of communicative visualizations, specifying communicative intent during design is ad hoc. Whether we are selecting from a set of visualizations, commissioning someone to produce them, or creating them ourselves, an effective way of specifying intent can help guide this process. Ideally, we would have a concise and shared specification language. In previous work, we have argued that communicative intents can be viewed as a learning/assessment problem (i.e., what should the reader learn and what test should they do well on). Learning-based specification formats are linked (e.g., assessments are derived from objectives) but some may more effectively specify communicative intent. Through a large-scale experiment, we studied three specification types: learning objectives, insights, and assessments. Participants, guided by one of these specifications, rated their preferences for a set of visualization designs. Then, we evaluated the set of visualization designs to assess which specification led participants to prefer the most effective visualizations. We find that while all specification types have benefits over no-specification, each format has its own advantages. Our results show that learning objective-based specifications helped participants the most in visualization selection. We also identify situations in which specifications may be insufficient and assessments are vital.

%% file: sections/01_introduction.tex
\section{Introduction}

Communicative visualizations are omnipresent. They exist in everything from news articles to scientific papers and from Web pages to television broadcasts. The people involved in the visualization design process must make design decisions based on their intents or goals. Unfortunately, most design guidelines do not connect communicative intent to the actual design. Instead, there is a significant amount of information on what makes communicative visualizations effective perceptually, focusing on making the visualization readable. However, this low-level information often fails to account for what kind of impact people would ultimately like to make: to influence the viewer cognitively (i.e., to have them learn something) or affectively (e.g., to have them believe something)~\cite{adar_communicative_2020,hohman2020communicating}.

Current design literature is often biased towards low-level cognitive efficiency~\cite{cairo2012functional,few2004show,tufte2001visual,tufte2006beautiful,wong2013wall}. This literature is built on significant academic foundations~\cite{ware2019information} that have allowed us to identify which encodings can be read accurately and quickly. However, if designers only rely on cognitive efficiency evaluation methods, such as how fast a viewer can decode the visualization, they might fail to consider higher-level communication goals for what they want their reader to be able to know or do. In most instances, being able to read the visualization is only the first step and may not even be the most crucial~\cite{clark2010graphics,hullman_benefitting_2011}. A designer may also want the viewer to remember the message, take personal action, or generate hypotheses. If the reader only reads the visualization but does not take any of these next steps, it would not be considered a good design. In addition to focusing on low-level cognitive efficiency, identifying higher-level communicative goals would benefit everyone involved in the design process.

There may be several stakeholders who are responsible for creating a visualization, ranging from those who have an initial idea for it, to those who create it, to those who choose the final version to publish. It's possible that there are different people for each of these steps, in the case of a client, a contractor, and an executive editor, or it's possible that one person carries out all steps of this process from start to finish. At every step, each stakeholder would benefit from a formal language for communicative intent. For creating a visualization, a clear communicative intent specification can help the designer identify their own goals and evaluate whether their visualization achieves them.  \revis{Professional designers may already know how to identify a communicative intent. Even so, they could still benefit from a framework to communicate with other stakeholders.} For a client commissioning a designer, a shared language of communicative intent would be helpful for the client to be able to succinctly express their goals to the designer. Sometimes, there might be several visualizations, either created by a designer or automatically generated by a computer, that a client would need to choose from in order to best fulfill their communicative goals. In this scenario, a formal structure for communicative intent can help the client analyze the different visualization designs to choose the best one. While there might be several stakeholders involved in creating, choosing, and publishing a visualization, all of them would benefit from a formal language for communicative intent to guide their design decisions, communicate with each other, and decide between multiple visualization designs. 

The framing of communicative visualization design as a learning design problem~\cite{adar_communicative_2020} is a formal structure that prioritizes higher-level intents. The stakeholder takes on the role of a teacher, explicitly defining their goals of what they want the viewer (student) to be able to do or remember after viewing the visualization. For example, \textit{the viewer will be able to recall the change in unemployment last year}, or \textit{the viewer will be able to determine the optimal treatment}. These learning objective statements are based on Bloom's Taxonomy~\cite{anderson_taxonomy_2004}, though many other variants are possible~\cite{biggs2014evaluating,wiggins2005understanding}. Critically, a \textit{specification} of a communicative visualization as a learning problem can lead to appropriate \textit{assessments}. A stakeholder can implement an actual test (e.g., a multiple-choice question ``post-test'') to determine if the visualization meets the specification criteria and satisfies the communicative intents. 

Even in the context of learning, there are many ways that a specification can be crafted for a communicative visualization. For example, \textit{learning objective statements} (e.g., \textit{the viewer will be able to describe the reasons why Norway is a Winter Olympics powerhouse}) are broad. This type of specification is high-level; it includes both the cognitive ability (e.g., the verb \textit{describe}) and knowledge (e.g., the `noun' \textit{Norway's success in the Olympics}). To achieve this, the viewer may need to read many facts, or \textit{insights}, that are in one or more visualizations. An alternative specification format might emphasize one or more of these lower-level insights: \textit{the high number of Olympic medals relative to population} or \textit{the many skating rinks that Norway has}. While insights may not be as broad as learning objectives, they may be easier to map to `graphical facts' or annotations and might be more comfortable for a designer to work with. A third specification format might privilege the assessments themselves--the tests we could use to validate learning: \textit{In the last ten Winter Olympics, how many times did Norway win the most medals?} While specific, this format may help the designer select a visualization that will do best in testing. Conversely, an over-focus on specific test questions or insights may lead to visualization choices that are not effective broadly. This is the design equivalent of `teaching to the test.'

The formats of these specifications are often transformable. For example, we could create one or more assessments for each insight or learning objective. Similarly, we can identify the insights that need to be communicated and learned to achieve a higher-level learning objective. We hypothesize that different specification forms emphasize different ways of thinking. Given these different specifications, designers or editors might make different choices. More critically, some of these choices might be better for learning outcomes. 

In this paper, we measure the impact of different specifications on visualization design choices. We contrast the visualization preferences when participants are exposed to different specifications in the form of learning objectives, insights, and assessments, and contrast these to baseline preferences. Additionally, we run a set of assessments against the possible visualizations to determine which produce the most accurate answers. Using both the preference data (i.e., the most preferred visualization given the specification) and accuracy (i.e., the best \textit{performing} visualizations), we identify when different specifications lead to better learning outcomes. Specifically, we find:
\begin{enumerate}[noitemsep,nolistsep]
    \item Adding specification information changes preferences over baseline judgments.
    \item Learning objectives and insight specifications guide designers to prefer better-performing visualizations.
    \item Evaluations of assessments with user testing provide critical feedback of what designs are effective.
\end{enumerate}

%% file: sections/02_related.tex
\section{Related Work }
We briefly examine various ways to describe visualization intent. We focus on summarizing the pros and cons of different approaches.

\subsection{Cognitive Effectiveness}
    A common approach in understanding communicative visualizations is a focus on the outcome rather than intent. Specifically, evaluations provide a mechanism to judge whether a visualization achieves a designated communication goal. 
    However, these approaches often emphasize cognitive effectiveness metrics. For example, one common way to evaluate a visualization is to test how fast and accurately a user can perform a task while viewing the visualization. Because designers can use low-level perceptual task taxonomies (e.g., as described by \cite{amar_low-level_2005}) at the beginning of their design process, these evaluation mechanisms can encompass designers' intent. Using a combination of tasks, designers can identify the visualization that is most effective for each task \cite{saket2018task}. 

    However, while benchmark usability tests are standard, these tasks are elementary and not representative of more complicated interactions. Furthermore, these measures may involve trade-offs between speed and accuracy \cite{north_toward_2006}. These generic and simplified tasks cannot evaluate more domain-specific and higher-level goals. This limitation inspires us to identify a better mechanism for goal specification and evaluation.
    

\subsection{Memorability}
        
 There are various memory evaluations for visualizations. For example, we could simply ask: ``if shown a visualization twice, would a viewer recall seeing it before?'' An alternative test may target a subgoal of the visualization by asking whether the viewer remembers a particular takeaway after the viewer has stopped looking at the visualization. Memorability is an essential part of communicative intent; designers should not be happy if the information went in one ear and out the other. Works in semiology such as \cite{bertin1983semiology,KindlmannAlgebraicVisDesign2014,mackinlay_automating_1986} offer insights into how meaning is formed and communicated through images and that they should be `memorizable.' Although this group of works has limited analyses for communicative visualizations, they argue that designers should consider long-term engagement and retention as valuable goals. 

Memorability tests shape designers' design decisions and direction. For example, visual images such as chart junk may increase the memorability of a visualization \cite{borkin_beyond_2016}.  Designers might also choose to add visual difficulties to a visualization to increase engagement, higher processing, and long-term retention \cite{hullman_benefitting_2011}. Our experimental design targets memorability as a key feature.
        
\subsection{Usability Tests and Qualitative Evaluations}
    Designers can evaluate visualizations with various usability tests. Specific techniques range from usability inspection metrics (as described by \cite{hollingsed_usability_2007}), running interface-focused user studies \cite{hollingsed_usability_2007}, and running qualitative measures such as interviews and observations \cite{isenberg_grounded_2008}. Evaluations of visualizations can be based on common design guidelines and heuristics, such as ``this visualization makes important information visually salient'' \cite{forsell_heuristic_2010, hearst_evaluating_2016}. For narrative visualizations, \cite{hogan2016elicitation} and \cite{nowak2018micro} discuss the potentials of using the Elicitation Interview technique to evaluate static and narrative visualizations. \revis{For visual analytic systems, process data could be collected as the user explores the data and can be used to evaluate the system \cite{chang2010learning}.}

    Compared to generic task taxonomies, usability tests are specific to the visualization systems and their communicative intent. Therefore, they have the potential to target more complex cognitive levels. Nevertheless, they can be costly and time-consuming. Based on heuristics used in the tests, these evaluations might require trained experts. 
    
\subsection{Storytelling, Draw Your Own, and Games}
    While most of the evaluation mechanisms target generic visualizations, some strategies help designers achieve communicative visualization objectives. For example, using visualization rhetoric as an analytical framework for narrative visualization can help designers make engaging, layered visualizations that have clear interpretation priorities\cite{hullman_visualization_2011}. Draw-your-own style visualizations ask viewers to chart out their expectations before seeing the data, an action that supports learning \cite{kim2017explaining}. Furthermore, bringing gamification to visualization has the potential to improve engagement \cite{Diakopoulos20011playable}. Nevertheless, these strategies describe \textit{how} to achieve an objective rather than offering a language to specify those objectives.

\subsection{Specifications}
    Overall, existing evaluation methods rarely focus on the specific higher-level communicative goals. While there are strategies to improve communicative visualizations, designers might still find it challenging to describe their intent due to a lack of resources and guidelines. As a result, designers often need to evaluate communicative visualizations from the cognitive efficiency perspective. 
    \revis{A different approach is to frame the evaluation of visualizations as a learning-based problem, both for communicative visualizations~\cite{adar_communicative_2020} and visual analytic systems~\cite{chang2010learning}.}
    Specifically, \revis{this framework} 
    connects studies in learning science and communication theory to construct an insight-to-learning-objective taxonomy that models designers' intent. 

    In the visualization community, insights are commonly thought of as the goal of visualizations \cite{card1999readings, yi2008}. A widely accepted definition of insight is a unit of information or knowledge that a visualization communicates \cite{chang_defining_2009}. An insight could be an observation of a fact, pattern, or relationship in the data \cite{chen2009toward}. Insights can range from simple to complex on several characteristics: complex, deep, qualitative, unexpected, and relevant \cite{north_toward_2006}. From a communicative viewpoint, designers could consider insights as the conclusion or the primary takeaway from the graph.
        
    With insights as communicative intent, designers aim to guide the viewer to understand potentially multiple data insights. Before or during the design process, designers will know what specific insight(s) they want to communicate. There are several limitations to this type of specification. Using insights to formulate communicative intent focuses on the data, therefore, often does not capture how the \textit{viewer} will interact with the data. Also, insights might not capture all of the different kinds of communicative goals, such as outcomes or actions that designers want the viewer to do after seeing the visualization.

    An alternative to insight is learning objectives. We adapt learning objectives from Bloom's Taxonomy for describing communicative intent for visualizations \cite{adar_communicative_2020}. Through the lens of a teaching problem, designers take on a teacher's role, explicitly defining their goals of what they want the viewer (student) to do or remember after viewing the visualization. Learning objectives have a structure of \textit{The viewer will [verb] [noun]}, where verbs specify cognitive change and the noun is a piece of knowledge. Bloom's taxonomy \cite{anderson_taxonomy_2004} provides a structure for designers to articulate what they want their audience to do effectively and at varying complexity levels. 

%% file: sections/03_intent.tex
\section{Visualization Goal Specification}
\label{intent}




In our study, we used three types of specifications to frame communicative intent: learning objectives, insights, and assessments. Each type reflects a different way of considering how the viewer might learn. 

\begin{figure}[t]
 \centering 
 \includegraphics[width=\linewidth]{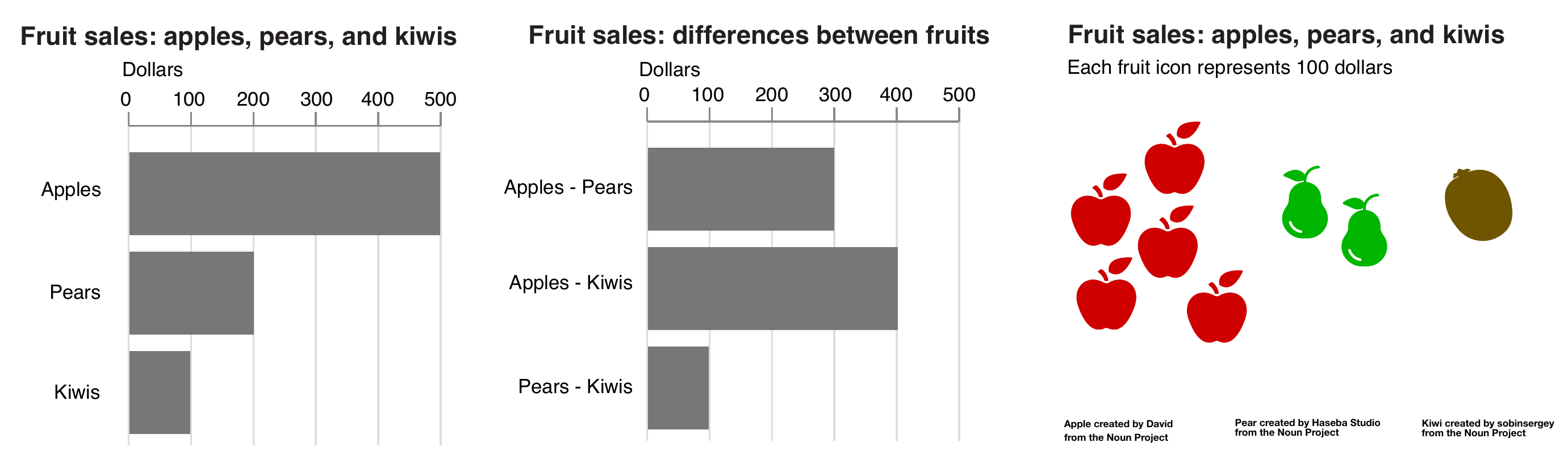}
 \caption{Example visualizations for our fruit dataset.}
 \label{fig:spec_example}
\end{figure}

To briefly introduce these types of specifications, we discuss a simple example of a dataset with three values of fruit sales: \$500 in apple sales, \$200 in pear sales, and \$100 in kiwi sales. This data can be visualized as a bar chart with three groups (Figure~\ref{fig:spec_example}, left). For this dataset, we can imagine that the person creating or choosing a visualization has learning objectives: \textit{$LO_1$: the viewer will recall the amount sold of each fruit} and/or \textit{$LO_2$: the viewer will recall the differences between fruit sales}. Depending on the learning objective(s) and their relative importance (and factoring in the properties of the audience), we might make different choices when selecting one of the visualizations in Figure~\ref{fig:spec_example}. Learning objectives can build off each other (e.g., we have to recall something before we can use it for synthesis tasks). Mathematically, learning objectives may also be equivalent or overlapping. In our simple example, one can calculate the differences between sales from the values. Within our two learning objective, there are six different potential \textit{insights} that arise:

\begin{itemize}[noitemsep,nolistsep]
    \item $I_1$: There was \$500 in apple sales [$LO_1$]
    \item $I_2$: There was \$200 in pear sales [$LO_1$]
    \item $I_3$: There was \$100 in kiwi sales [$LO_1$]
    \item $I_4$: Apple sales were \$300 more than pear sales [$LO_2$]
    \item $I_5$: Apple sales were \$400 more than kiwi sales [$LO_2$]
    \item $I_6$: Pears sales were \$100 more than kiwi sales [$LO_2$]
\end{itemize}

Depending on our learning goal, we may focus on the $LO_1$ or $LO_2$ insights. Note that our definition of insight centers on facts or statements about data. These need not be the `a-ha' style insights experienced in analytical visualization forms~\cite{chang_defining_2009}. Rather, these are communicative insights--those chosen by the visualization designer--that they would like the viewer to recognize as `facts' both during and after reading a visualization. We specifically relate insights to the learning objectives in that they are statistical statements about the data that emerge from the noun portion of the learning objective. In most situations, we expect that the viewer would be able to validate these insights as true or false. In our example above, there may be multiple insights that arise from one learning objective. Thus, we can specify our intent using these statements (or some representative sub-sample).

As we generated insights, we can also create \textit{assessments}. For $LO_2$, or equivalently, the second set of insights ($I_4, I_5, I_6$), we could write a `test' to evaluate whether the viewer had recalled the differences between fruit sales. There are multiple possible assessments that we could use. For example, we could ask the viewer to answer ``What's the difference between apple sales and kiwi sales?'' Alternatively, we could ask a true or false question, ``True or False: apple sales were \$200 more than kiwi sales?'' There are also easier, but insufficient assessments one could use, such as ranking fruits by sales. This question would assess whether the viewer knows the ordinal rank of each fruit, which would be a first step in knowing the exact differences. Another test would be to ask a question to assess $I_1$ ``What was the total value of apple sales?'' This assessment would give the stakeholder information on what fundamental information may be learned or lacking from the viewer. This assessment provides more detailed information on what parts of the visualization are falling short, allowing the designer to address any shortcomings. Just as we could create a specification of our intent using learning objectives, or insight statements, we could use a set of assessments instead. Critically, each form of specification has positive and negative features.

\begin{figure*}[t]
    \centering
    \includegraphics[width=0.9\textwidth]{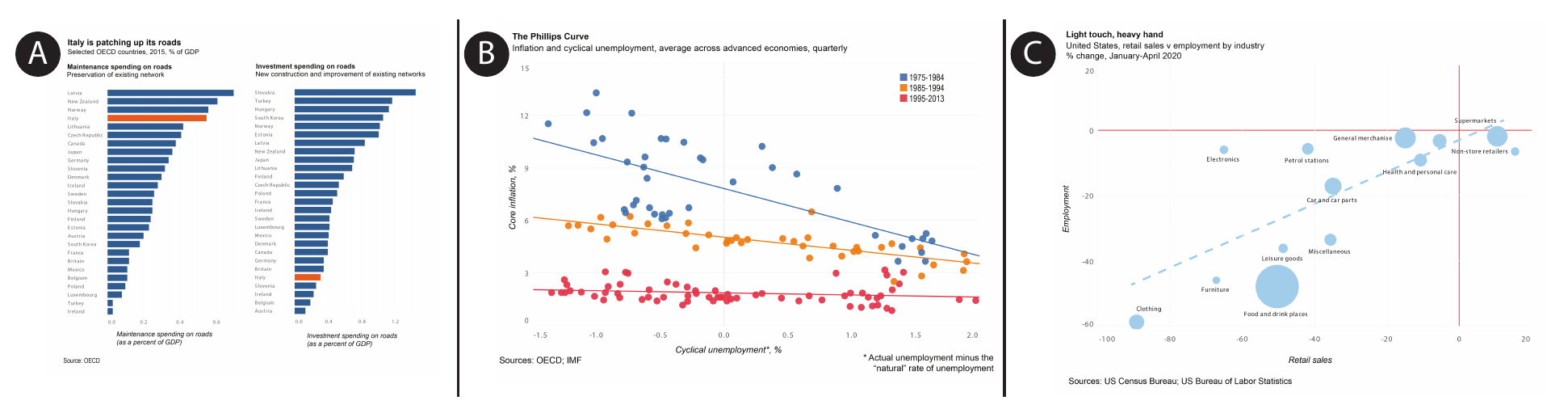}
    \caption{Recreations of the three Economist Graphic Detail visualizations: (A) Italian road quality~\cite{italy}, (B) Phillips curve~\cite{phillips}, and (C) COVID and employment~\cite{covid}.}
    \label{fig:origvis}
\end{figure*}

\subsection{Learning Objectives}
Our simple learning objective above took the form of, ``the viewer will recall the amount sold of each fruit. ($LO_1$)'' The template for this statement is derived from the application of the Bloom's Taxonomy~\cite{adar_communicative_2020,anderson_taxonomy_2004}. Learning objectives have a structure of \textit{the viewer will [verb] [noun].} Verbs correspond to one of the cognitive dimensions: \textit{remember}, \textit{understand}, \textit{apply}, \textit{analyze}, \textit{evaluate}, and \textit{create} (each with sub-categories and synonyms). These categories are largely hierarchical, with lower-level tasks (e.g., \textit{recall}) contrasting to high-level objectives (e.g., \textit{generate}).  Nouns, in our formulation, are specific. For example, ``unemployment levels in California'' or ``the best algorithm.'' Nouns also fall broadly into \textit{factual}
, \textit{conceptual}, \textit{procedural}, and \textit{metacognitive} knowledge. Taken together, a more sophisticated learning objective might be:

\begin{quoting}
$LO_3$: The viewer will summarize the relationship between loss of employee and loss in sales across different industries due to COVID-19.
\end{quoting}

We have previously demonstrated how communicative visualization intents can be mapped to this taxonomy~\cite{adar_communicative_2020}. Our work demonstrated that designers can map many of their intentions to this framework. The benefit of this approach is that learning objectives can encapsulate a wide range of lower-level goals (e.g., remembering a set of facts) through succinct high-level descriptions. However, creating objectives is difficult even with training~\cite{gronlund2008gronlund,raible2016writing}. Thus, constructing objectives as specifications may not always be practical or convenient.

\subsection{Insights}
An alternative way of considering learning is to ignore the cognitive dimension and focus on key data ``facts'' or ``insights.''  Though there are numerous ways of modeling insight~\cite{chang_defining_2009}, we are concerned with a narrower definition. In the context of communicative visualization, insights are statements about data that the designer would like the viewer to recognize and internalize. \revis{In our fruit sales example, we suggested an insight ($I_1$) of, ``There was \$500 in apple sales.''} An insight may be the ``noun'' portion of a learning objective but is often more atomic and more specific. For example, if we consider the objective ($LO_3$) above, an insight form might be: 
\begin{quoting}
    $I_7$: There is a linear relationship between loss of employees and loss in sales across different industries due to COVID-19.
\end{quoting}

Communicative insights are often easy to identify and articulate. Many visualization annotations are reflections of insights in graphical form. Thus, a visualization goal specification may consist of one or more insights. The designer can focus on this set when designing or selecting a visualization. However, because insights are low-level and speak directly to data facts, we may require many insights in our specification to cover one learning objective. Too many insights in a specification may overwhelm the designer. Too few, and they might fixate on a narrower objective than intended. Additionally, because of their data orientation, insights do not always support other types of learning. For example, we may want a communicative visualization that teaches the viewer how to apply an algorithm (e.g., a triage flowchart at a hospital). It is more difficult to formulate that type of goal as an insight statement. 

\subsection{Assessments}
Assessments are an alternative, outcome-focused way of describing learning. A feature of learning objectives is an extensive mechanism for mapping objectives to concrete assessments~\cite{haladyna2012developing,popham2003test}. For example, an assessment for the objective/insight ($LO_3/I_7$) above might be: 
\begin{quoting}
Q: What is the relationship between employment and sales? \\ A: There exists a linear relationship where a loss of sales results in a loss of employment.
\end{quoting}
Even multiple-choice style questions can be combined to identify if learning has been achieved. When these are insufficient, open-ended questions with a rubric may also work. In our context, assessments of this type are used to measure ``program effects.'' That is, we are not concerned with a single viewer's performance, but rather the effect of the visualization on all the viewers. By specifying visualization goals through a list of questions, the designer may be forced to reckon with their choice relative to a very specific set of outcomes. The designer should create or select the visualization that will lead to the best assessment results. An assessment-based formulation may have both the benefit and disadvantages of focusing the designer on an outcome metric. Instructors will recognize the danger here of, ``teaching to the test'' (or in this case, ``designing to the test.''). As with insights, we may need many assessment questions to describe one learning objective or `triangulate' if learning has happened.

To summarize, we note that these three specification formats are connected to each other, but vary in systematic ways. Learning objectives, which are explicit in specifying the cognitive impact, are often broader and more abstract than insights or assessments. Assessments are more specific than insights, focusing designers on only one possible question they could ask of a viewer. We expect that these differences in specification format will lead to differences in designer preferences.  It is also ultimately possible that the best specification will be a mix of these approaches. The different forms of specification have different levels of granularity, ambiguity, generalizable properties, communicative efficiency, and other features that would make their use more or less desirable for describing intent. While these facets are all potentially interesting, we begin with a more high-level question: does communicative intent specified in one of these forms lead to better learning outcomes?

%% file: sections/04_study.tex
\section{Study Design}



In our study, we conduct two experiments to investigate whether the type of specification had an impact on preferences for designs and, ultimately, their effectiveness. More specifically, our main research questions are:
\begin{enumerate}[noitemsep]
    \item{Do different specifications (learning objectives, insights, assessments) affect the choice of visualization designs?} 
    \item{Are the chosen designs more effective than the alternatives?}
\end{enumerate}


\subsection{Materials}




\begin{figure*}[t]
 \centering 
 \includegraphics[width=\textwidth]{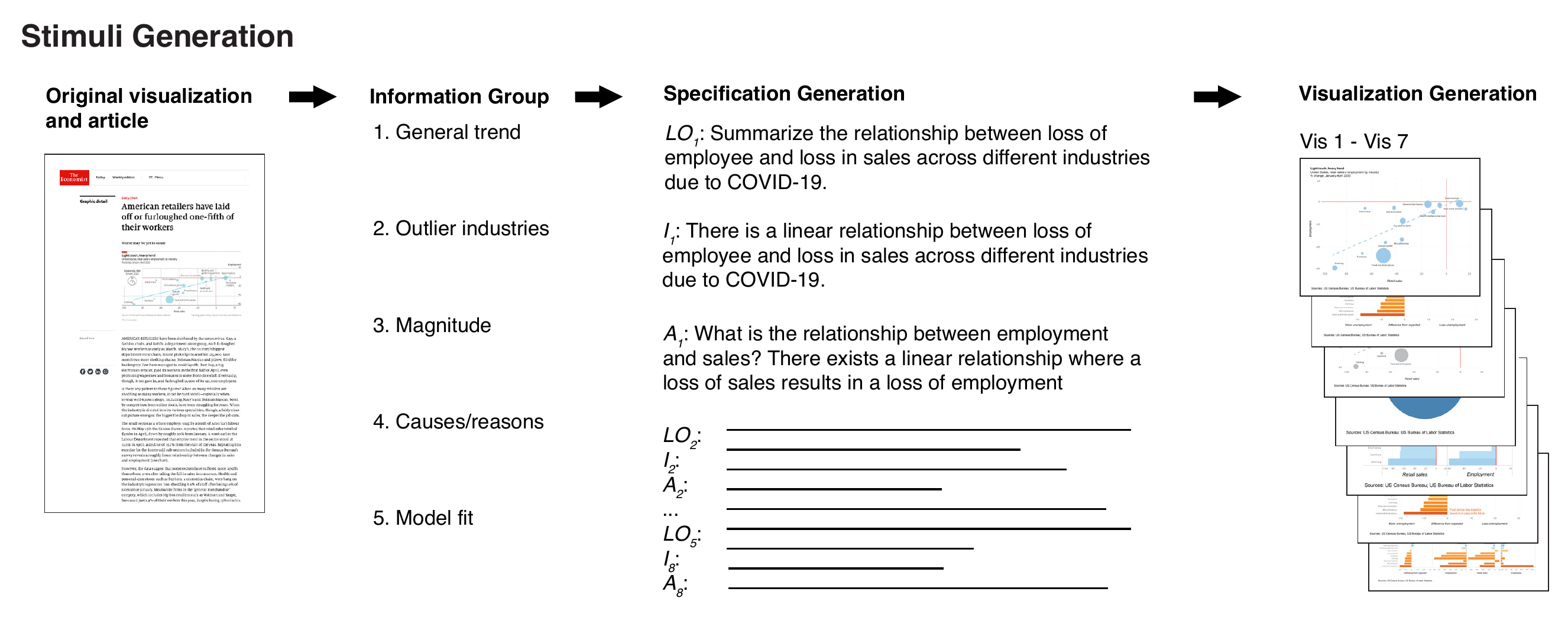}
 \caption{A diagram of the specification and visualization generation. The three specifications shown stem from information group 1. The top visualization was created specifically for information group 1. A full set of the specifications and visualizations are in Supplemental Materials. }
 \label{fig:stimuli_diagram}
\end{figure*} 

To create materials for this experiment, we identified a set of three starting articles from which we generated visualization designs, learning objectives, insights, and assessments. 
We used three articles from The Economist's Graphic Detail series: ``American retailers have laid off or furloughed one-fifth of their workers''~(COVID)~\cite{covid}, ``The Phillips curve may be broken for good''~(Phillips Curve)~\cite{phillips}, and ``Italy spends lots fixing old roads, not enough building new ones''~(Road Quality)~\cite{italy}. Each of the articles is centered on one or two visualizations with a relatively short article that describes key details (see Figure~\ref{fig:origvis}). Because the text is mostly anchored to elements of the visualization, it becomes easier to infer the likely communicative intents and reverse engineer plausible specifications. 

For each key fact discussed in the article, we developed a learning objective, the same information expressed as an insight(s), and the assessment(s). These linked \textit{information groups} of specifications all focus on communicating the same element of the visualization. By having roughly equivalent specifications, we can meaningfully compare specification types to each other in our study. Because of the lack of resources on how to formulate learning objectives and how they relate to insights and assessments, we explored possible ways to make linked groups of specifications. We ultimately decided to create specification groups by having learning objectives encompassing one or more insights, and each insight has one corresponding assessment question. \revis{Because of this structure, we ended up with fewer learning objectives than insights and assessments---one learning objective could comprise of more than one insight and assessment.}

For the COVID retail article~\cite{covid}, a broad learning objective is ``Summarize outlier industries/ differences in sectors.'' The three insights in this information group are: ``Health and Personal care have an expected amount of layoffs''; ``Food service has more layoffs than expected''; and ``Electronic stores have fewer layoffs than expected.'' However, in more `basic' (i.e., simpler) learning objectives, the information between the two specifications are more similar. For example, one of our learning objectives was ``Recall that American's retailers have laid off of furloughed a fifth of employees because of COVID-19,'' and the single corresponding insight was ``American's retailers have laid off of furloughed a fifth of employees because of COVID-19''. 
To constrain our experiment, we created one assessment per insight. In reality, one could construct assessments that cover multiple insights simultaneously or, conversely, multiple assessments for the same insight. 

After creating the set of linked information groups of specifications, we created several different visualization designs. The original visualizations from the Economist are professionally constructed (our reconstructed versions are shown in Figure~\ref{fig:origvis}). These visualizations provided us with a starting point for alternatives. With learning in mind, we generated multiple visualizations of varying expressiveness~\cite{mackinlay_automating_1986}. We created designs that we thought would be the most effective for a particular specification. For example, we designed a simple pie chart highlighting 20\% for the specification ``American's retailers have laid off or furloughed a fifth of employees because of COVID-19.''


Figure~\ref{fig:stimuli_diagram} illustrates the process that we used to generate the materials. Table~\ref{materials} summarizes the materials generated for the three articles. The complete instrument is available in our supplemental materials.

\begin{center}
\begin{table}[t]
\begin{tabular}{ r c c c c }
 \textbf{Dataset} & \textbf{Vis.} & \textbf{Learn. Obj.} & \textbf{Insights} & \textbf{Questions} \\
 \hline
 COVID & 7 & 5 & 8 & 8 \\ 
 Phillips Curve & 8 & 7 & 9 & 9 \\  
 Road Quality & 4 & 3 & 4 & 4 \\
\end{tabular}
\caption{\label{materials} Materials produced for our experiments. Each learning objective include one or more insights and each insight is associated with one multiple choice assessment question.}
\end{table}
\end{center}






\subsection{Preference Experiment}
In our first experiment, we collected participant preferences for multiple visualization designs. Though we only had our participants choose between different visualization designs, our study may be more broadly applicable to designers that actually create different designs, weighing different design decisions based on communicative goals. 

Participants in our study began by reading an informed consent document and completing two qualification questions. These questions tested the participants' abilities to correctly extract information from a bar graph and a scatterplot. This ensured a level of visual literacy. After passing the qualification questions, participants read instructions and completed three practice questions with feedback. \revis{In this experiment, the visualizations were shown within-subjects and the specifications were shown between-subjects.} Participants previewed all of the visualizations from one dataset on the next page. This preview was to encourage participants to compare the visualizations to each other in their ratings. On the following page, we showed each visualization with a specification and elicited a rating on a five-point Likert scale from ``Extremely good'' to ``Extremely bad.'' Each participant rated each visualization once based on a single specification. 

In previous pilot iterations, we experimented with having participants rank graphs either in a pair or in a group. The advantage of rating each graph in a Likert scale instead of ranking is that we could evaluate the magnitude of differences between the graphs. Additionally, rating all of the visualizations meant that preferences could be relative to all graphs in the set. 

To obtain baseline ratings of visualizations, we conducted the same experiment \textit{without} providing any specification. \revis{In this condition, we simply asked ``Please evaluate each visualization.''  Participants were free to interpret this as they chose.} With the same procedure as before, participants rated the visualization on its own without a specification. With this measure, we can adjust each preference based on how much it deviates from its baseline rating. This allows us to compare the visualizations to each other by adjusting for each graph's varying aesthetics and designs.

For the preference experiment, we hosted our experiment on Qualtrics and recruited 667 participants from Amazon Mechanical Turk. We compensated participants with \$1.00, and the median completion time was 6.2 minutes. For the baseline experiment of evaluating the visualizations without any specifications, we recruited 69 participants. We compensated participants with \$0.45, and the median completion time was 3.4 minutes. 

\subsection{\revis{Accuracy Experiment}}
\label{eval_description}
In contrast to the participant who decides which visualization is best (the message sender), we refer to the recipient as the \textit{viewer}. 
To compare the participant preferences to an empirical evaluation of the effectiveness of the design, we conducted another experiment to evaluate \textit{viewer} accuracy on the assessments. In order to evaluate the visualizations, we implemented three versions of the accuracy experiment. \revis{In this experiment, the visualizations were shown between-subjects.} We asked participants to answer the question in one of three conditions: before they saw the visualization (no-visualization baseline), while they were viewing the visualization (readability), or after they had seen the visualization and it was taken away (memorability). Each participant only viewed one test question and one visualization.
The no-visualization baseline condition allows us to determine the question's baseline difficulty based on guessing and prior knowledge. Then, we evaluated readability accuracy for answering the assessment while viewing the visualization. Finally, we evaluated the memorability accuracy on how effective the visualizations were for information that was to be remembered. 

For the accuracy experiment we again used Qualtrics and recruited participants from Amazon Mechanical Turk. We recruited 203 participants for the no-visualization baseline assessment (\$0.10, median completion time = 1.1 minutes), 1658 participants for the readability assessment during the visualization (\$0.10, median completion time = 1.4 minutes), and 1126 participants for the memorability assessment (\$0.20, median completion time = 2.1 minutes).

\begin{figure}[t]
 \centering 
 \includegraphics[width=.9\linewidth]{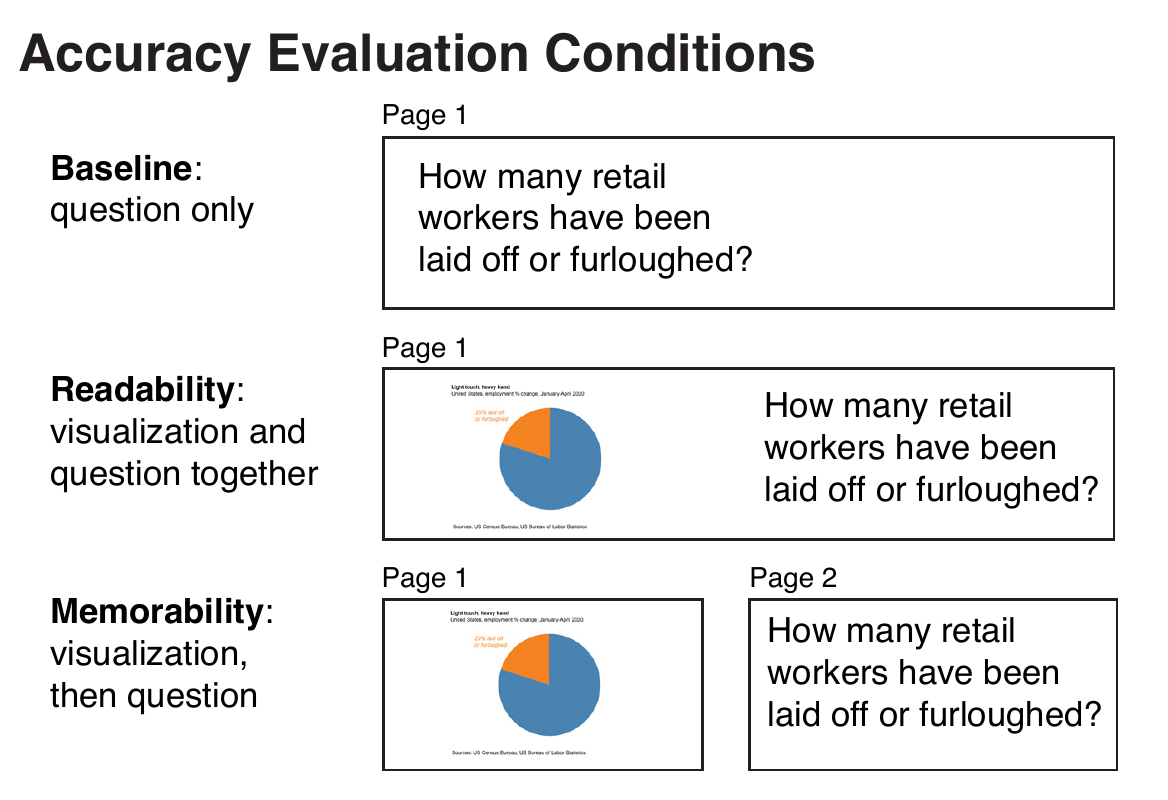}
 \caption{Three conditions: baseline, readability, and memorability.}
 \label{fig:eval_diagram}
\end{figure}

%% file: sections/05_result.tex
\section{Data Analysis}


After collecting the experimental data, we are able to return to our two main research questions. First, we investigate how different visualization specifications (learning objectives, insights, assessments) impact the choice of visualization designs. Then, we determine if the chosen visualizations are more effective than the alternatives.



\subsection{Preferences}
Our analysis begins by solely looking at preferences. That is, we explore whether the different types of specifications changed preference ratings of visualizations. Recall that participants rated their preference of visualizations on a five-point Likert scale from ``Extremely good'' (5) to ``Extremely bad'' (1). In our analysis, we look at both the raw averages of the preference scores and the rank of visualizations in relation to each other.

\begin{figure}[t]
 \centering 
 \includegraphics[width=.8\linewidth]{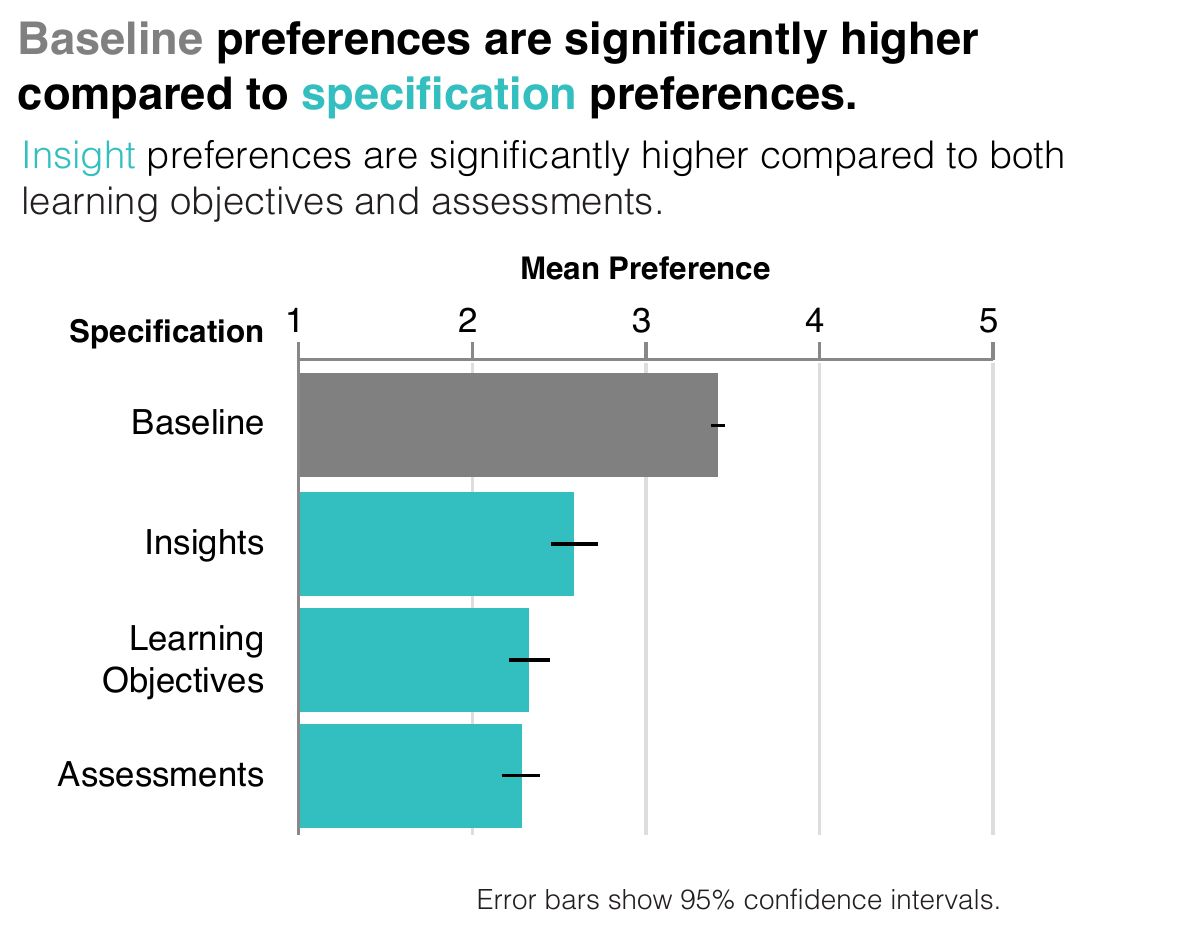}
 \caption{The average preference for each of the specification types: no specification (baseline), learning objectives, insights, and assessments. Error bars show 95\% confidence intervals. Participants preferred visualizations more when rating them with no information.}
 \label{fig:prefbars}
\end{figure}

First, we examined whether there were differences in average preference for the type of specification (learning objectives, insights, assessments, or no information). A Kruskal-Wallis rank sum test showed that preferences significantly differed by specification type, $H (3) = 74.35, p < 0.001$. Post-hoc Dunn's test with a Bonferroni correction showed that preferences for the three specifications were significantly lower than the baseline of no information, $p < 0.001$. On average, the participants rated the visualization lower when evaluating based on a specification compared to no specification (see Figure \ref{fig:prefbars}). A possible explanation for this is that the constraints imposed by specifications lead participants to be more critical of the visualizations. Additionally, preference scores from the insight specification were significantly higher than the learning objectives specifications ($p = 0.03$) and the assessment specifications ($p = 0.02$). 

Next, we looked at the similarity of the ranks of preferences. Even though the baseline preferences are significantly higher than the specification preferences, the visualizations may all be ranked the same. We conducted a Spearman correlation between the \textit{ranks} of the visualizations for each specification type. We found that all specification types were correlated with each other. Table \ref{tab:pref_rank} shows correlations of preference ranks for the specification types. Thus, preference ranks are positively correlated with each other, though clearly this is imperfect. 


\begin{table}[h]
    \centering
    \begin{tabular}{r c c c c }
        & \textbf{Learning Obj.} &  \textbf{Insight} & \textbf{Assessment}  \\
        \textbf{Insight} & 0.49 & - \\
        \textbf{Assessment} & 0.49 & 0.48 & - \\
        \textbf{Baseline} & 0.57 & 0.67 & 0.48 \\ 
    \end{tabular}
    \caption{Spearman correlation coefficients of preference ranks for each of the specification types: learning objectives, insights, assessments, and no specification (baseline). All correlations are significant ($p < 0.001$). Preference ranks are positively correlated with each other.}
    \label{tab:pref_rank}
\end{table}


We investigated whether different specification types lead participants to have different `most-preferred' visualizations. For each specification group, we looked at the most-preferred visualization for each specification type. Out of the 21 groups, four groups had different most-preferred visualizations for all three specifications. Eleven groups had one specification lead to a different most-preferred visualization, while the other two specifications lead to the same most-preferred visualization. Six groups had all three specifications lead to the same most-preferred visualization. 
See Figure~\ref{fig:teaser} for an example of an information group where each specification type leads to a different most-preferred visualization. 

Additionally, we can contrast the baseline most-preferred visualization (i.e., the one chosen based on considerations without specifications) against the specification's most-preferred visualization. We find in six cases the baseline matched no specifications (i.e., given a specification, the visualization was never most-preferred), seven times it matched one specification, five times it matched two specifications, and three times it matched all three specifications. This demonstrates the potential pitfall in a completely specification-free choice.

Just as we study the most-preferred visualization, we can also consider the least-preferred one. Out of the 21 groups, five groups had different least-preferred visualizations for all three specifications. Seven groups had one specification lead to a different least-preferred visualization, while the other two specifications lead to the same least-preferred visualization. Nine groups had all three specifications lead to the same least-preferred visualization. This indicates more agreement on the least-preferred choice given specifications. When considering the least-preferred visualization from the no-specification baseline, we find a similar pattern. In five cases the least-preferred no-specification visualization did not match any of the least-preferred visualizations given a specification. That is, the specification caused participants to reassess the `worst' visualizations. In six cases the least-preferred no-specification matched one specification, five times it matched two specifications. In five cases it matched all three specifications. In those few cases, the initial `gut' preference on the worst visualization corresponded to the preference after observing the specification.

To summarize, specifications affected visualization preferences. When given no information about the communicative goal, participants rated the visualizations higher than when they were given a specification. Out of the three specification formats, participants rated visualizations higher when viewing insights compared to either learning objectives or assessments. This suggests that insights may have more of a direct mapping onto graphical features of the visualization. Even though there was a difference between preferences, we found that the ranks of designs among specification types were correlated with each other, indicating some similarities between formats. 


\subsection{Accuracy}
While understanding preferences is interesting, we would ultimately like to know if specification formats lead participants to select more effective visualizations. We do so by analyzing the viewer accuracy on the assessments. We investigate three evaluative conditions: baseline, readability, and memorability (see Figure~\ref{fig:eval_diagram}). Recall that baseline is the performance \textit{before} seeing the visualization, readability is the performance \textit{while} seeing the visualization, and recall is \textit{after} seeing it \revis{(as described in section \ref{eval_description})}.

\begin{figure}[t]
 \centering 
 \includegraphics[width=.8\linewidth]{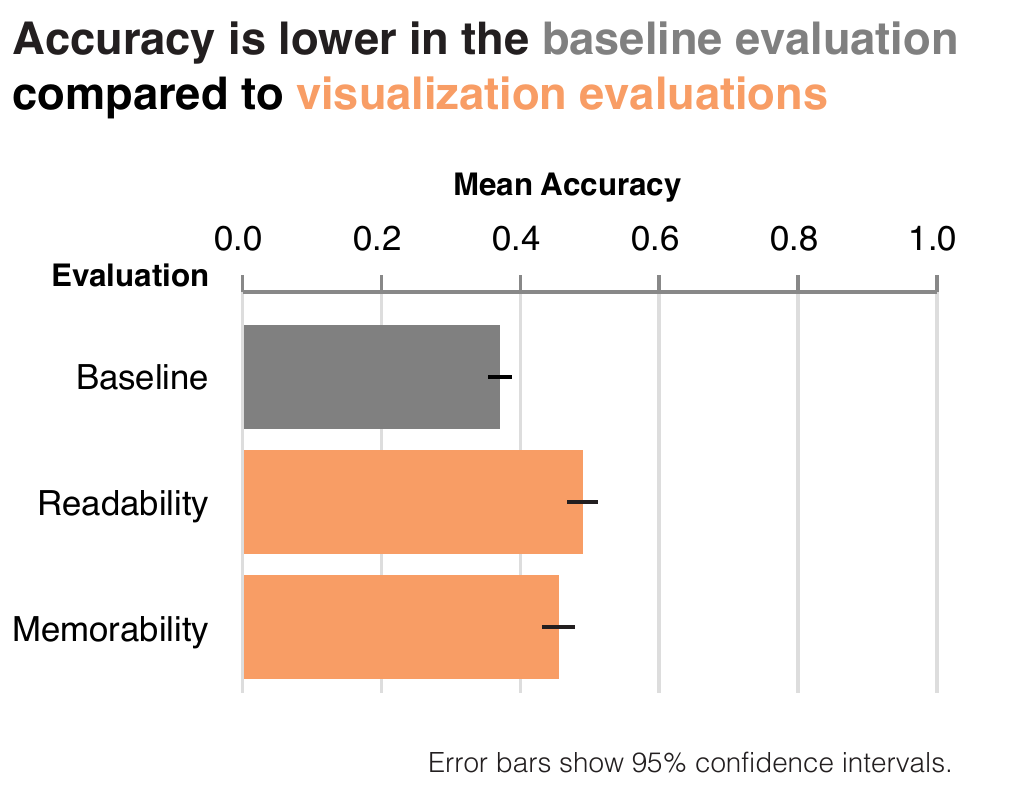}
 \caption{The average accuracy for each of the conditions: no-visualization baseline, readability, and memorability. Error bars show 95\% confidence intervals. Participants performed equally as good on the memorability assessments as they did on the readability assessments.}
 \label{fig:evalbars}
\end{figure}

We start by examining if there were differences in average accuracy between the three conditions (baseline, readability, and memorability). 
An ANOVA showed that accuracy significantly differed by condition, $F (2, 429) = 10.57, p < 0.001$. Post-hoc tests with a Bonferroni correction showed that baseline accuracy is significantly lower than the readability accuracy ($p < 0.001$) and memorability accuracy ($p = 0.005$). Memorability accuracy is not significantly different from readability accuracy ($p = 0.6$). 
Overall, the baseline accuracy for questions was approximately chance ($M = 0.37, SD = 0.18$). As expected, readability accuracy ($M = 0.49, SD = 0.24$) increased from the baseline of answering questions without the visualization. The memorability accuracy was between those two conditions ($M = 0.45, SD = 0.25$), though not significantly different from readability accuracy. These results are consistent with expectations (see Figure~\ref{fig:evalbars}). Without a visualization viewers perform the worst. While looking at the visualization viewers perform the best (in many cases they can decode the answer from the visualization). When we take the visualization away, viewer performance is slightly lower as viewers have to rely on their memory to answer the question (but not significantly different from when viewing the visualization). We note that performance in the memorability variant may decline as additional time or distractions are introduced.

While the general baseline-readability-memory pattern was consistent, we found that the baseline difficulty of the questions without a visualization varied. The baseline difficulty ranged from an accuracy of 0.11 (more likely to answer incorrectly than chance) to 0.80 (more likely to answer correctly than chance). We use the baseline difficulty to adjust questions to their underlying difficulty. We calculate the increase (or decrease) in accuracy from the baseline to the readability or memorability accuracy. This allows us to study the specific effect that the visualization has on accuracy.

\subsection{Preference and Accuracy}


We now compare the preferences to the accuracy to understand if the specification type influences participants to prefer better or worse visualizations. In this section, accuracy is defined as the increase (or decrease) in accuracy from the no-visualization baseline to the readability or memorability accuracy, as described above.

\begin{figure}[t]
 \centering 
 \includegraphics[width=\linewidth]{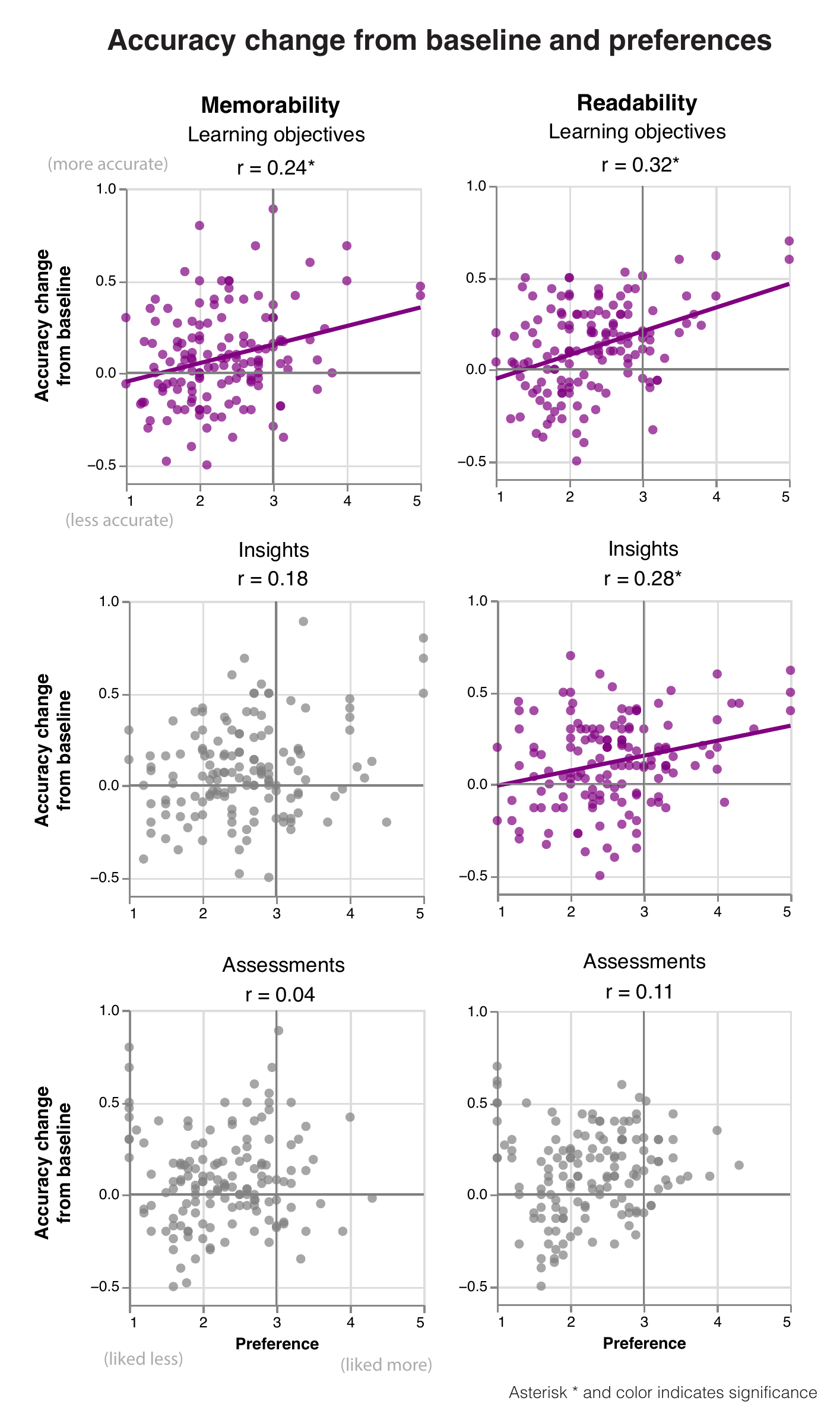}
 \caption{Scatterplots of preferences and accuracy change from baseline for the three specifications.}
\label{fig:scatterplot}
\end{figure}

We computed Spearman's correlation coefficient for each specification between preference and change in accuracy from the baseline for readability accuracy and memorability accuracy. There is a significant relationship between preferences for learning objectives and change in accuracy for both the readability accuracy and the memorability accuracy (see Table \ref{tab:pref_acc} and Figure \ref{fig:scatterplot}). For insights, there is a significant relationship between preferences and change in accuracy for readability accuracy, but not memorability accuracy. For assessments, there was not a significant relationship between preferences and change in accuracy for neither readability accuracy nor memorability accuracy. Additionally, there is not a significant relationship between baseline preferences and accuracy. When looking at the difference between the two accuracy conditions, we find the same trends for accuracy in the memorability condition as the readability condition, but not as strong. These correlations suggest that learning objective specifications lead people to prefer better visualization designs.

\begin{table}[t]
    \centering
    \begin{tabular}{r c c c c }
          &  \multicolumn{2}{c}{\textbf{Readability}} & \multicolumn{2}{c}{\textbf{Memorability}}\\
        \textbf{Specification} & \textbf{r} & \textbf{p} & \textbf{r} & \textbf{p}\\
        \hline
        Learning Objectives & 0.32 & $<0.001$* & 0.24 & 0.003* \\
        Insight & 0.28 & $<0.001$* & 0.18 & 0.03 \\
        Assessment & 0.11 & 0.19 & 0.04 & 0.65 \\
        Baseline & -0.02 & 0.85 & 0.07 & 0.41
    \end{tabular}
    \caption{Spearman's correlation coefficients between preferences and the adjusted accuracy for each specification. Adjusted accuracy is defined as change from the no-visualization baseline to either the readability assessments or the memorability assessments. P-values are judged at the bonferroni-adjusted significance level of p = 0.0125.}
    \label{tab:pref_acc}
\end{table}

A multiple regression model predicting the change in accuracy (baseline to readability) from baseline preference, learning objective preference, insight preference, and assessment preference was significant, F(4,139) = 7.80, R\textsuperscript{2} = 0.18, $p < 0.001$. Learning objective preferences and insight preferences significantly predict accuracy (see Table \ref{tab:regression}). Neither the assessment preferences nor the baseline preferences were significant predictors. Consistent with the correlation tests, we found similar results for a multiple regression model predicting change in accuracy for baseline accuracy to memorability accuracy, F(4,139) = 4.91, R\textsuperscript{2} = 0.12, $p < 0.001$.

\begin{table}[t]
    \centering
    \begin{tabular}{r c c c}
        \textbf{Readability} &  \textbf{b} & \textbf{Std. Err} & \textbf{p} \\
        \hline
        Learning Objectives & 0.1 & 0.03 & $<0.001$* \\
        Insight & 0.06 & 0.03 & 0.04* \\
        Assessment & -0.02 & 0.03 & 0.5 \\
        Baseline & -0.07 & 0.05 & 0.1\\
        \\
        \textbf{Memorability} &  \textbf{b} & \textbf{Std. Err} & \textbf{p} \\
        \hline
        Learning Objectives & 0.07 & 0.03 & 0.02* \\
        Insight & 0.07 & 0.03 & 0.02* \\
        Assessment & -0.05 & 0.03 & 0.2 \\
        Baseline & -0.008 & 0.05 & 0.9\\
    \end{tabular}
    \caption{Multiple regression models predicting accuracy from baseline preference, learning objective preference, insight preference, and assessment preference. The first model is predicting change from baseline accuracy to \textit{readability} accuracy; the second model is predicting change from baseline accuracy to \textit{memorability} accuracy.}
    \label{tab:regression}
\end{table}

\begin{figure*}[tbhp]
 \centering 
 \includegraphics[width=.9\textwidth]{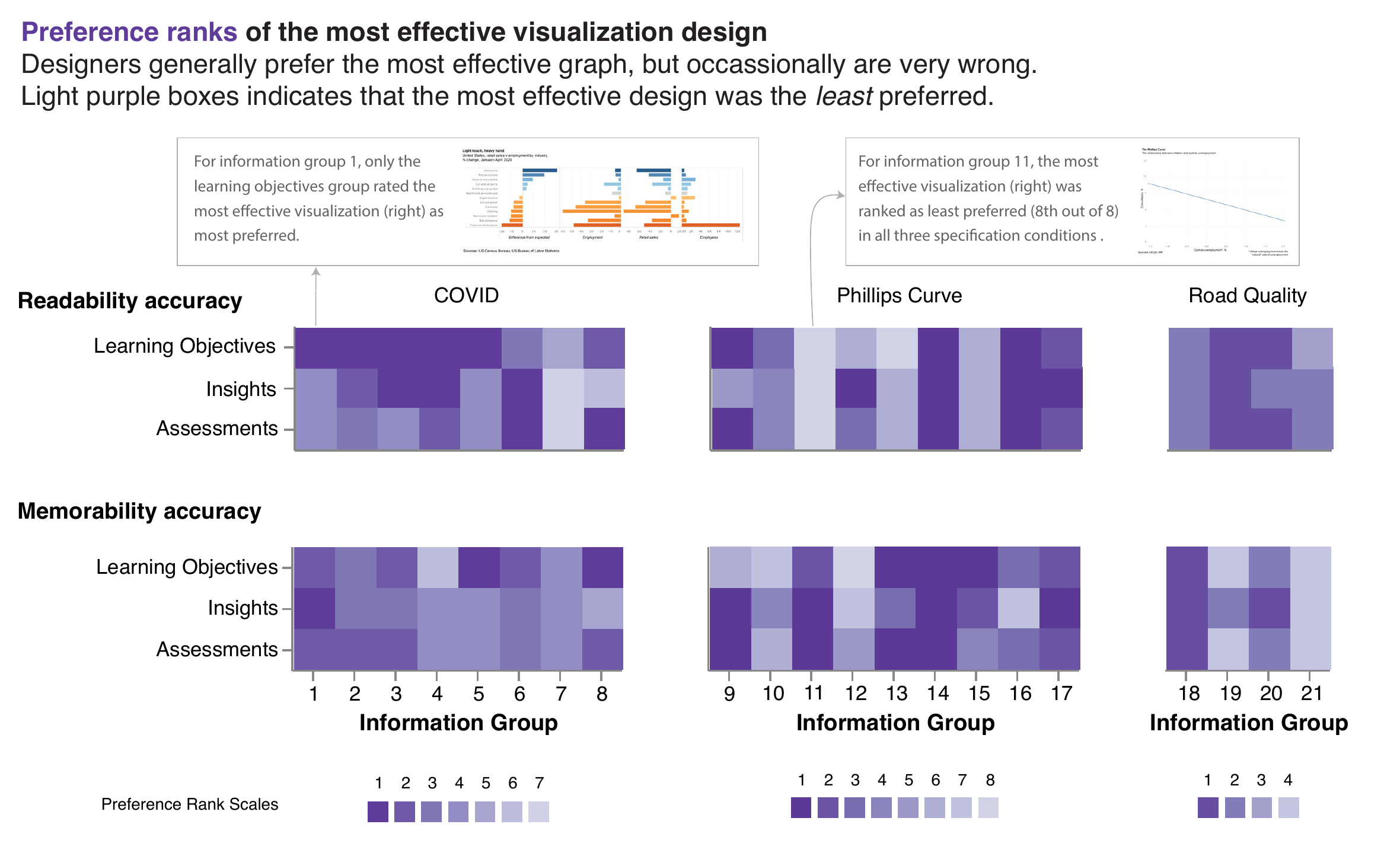}
 \caption{The visualization with the highest accuracy (for both readability assessments and memorability assessments) was identified for each information group. We show the rank of that visualization for each of the specifications (Learning objectives, Insights, Assessments). }
 \label{fig:best_pref_rank}
\end{figure*}

Furthermore, we found that designer preferences for graphs do not always line up with assessments. For each question, we identified the design that had the highest accuracy from the baseline and looked at preferences for that design. In a few cases, participants identified the graph with the highest accuracy as the \textit{least} preferred. For example, question 11 - ``Q: What is the Phillips Curve? A: The trade-off between unemployment and inflation'' is most correctly answered (in the readability condition) with a very minimal line chart. However, participants rated that chart as least preferred for all specifications (see Figure \ref{fig:best_pref_rank}). This shows that articulating a specification is important but sometimes insufficient. Thus, while assessments may not be the best specification format, they may be important in evaluating designer choices.

%% file: sections/06_discussion.tex
\section{Discussion}\label{discussion}

Our project investigated the effects of different specification types on selecting `better' visualizations. We found that just having a specification changed preferences over no information at all. More importantly, specifications led participants to prefer more effective visualizations, while baseline preferences did not. Our result shows that it is important to define a communicative intent as part of the design process. Specifying a communicative intent can guide the designer to focus on a goal, and help them make better design choices towards this goal.

In our analysis of the specifications, we found evidence that the three specification forms are similar, but may lead to significant differences in some situations. The informational content in each is similar, so it is not surprising that they were correlated with each other (Table \ref{tab:pref_acc}). However, the specification format can emphasize a different way of thinking about the goal. We found that preferences from learning objectives (and to a lesser extent, insights) were the most correlated with design effectiveness (Figure~\ref{fig:scatterplot}), suggesting they may be best overall. 

While our results conclude that specifications help people choose better visualizations, it also reveals that they may not be sufficient to choose the \textit{best} visualization. In some cases, the most effective designs for a specification may be intuitive. However, sometimes evaluations may reveal effective design options that would not have otherwise been considered. The analysis of the most effective design option illustrates this point (Figure~\ref{fig:best_pref_rank}). Participants generally preferred the most effective graph. However, occasionally they were very wrong, rating the most effective graph as their least preferred graph. Incorporating this evaluative feedback into the design process could help designers choose better graphs than preferences alone.

Our study also shows that having viewers complete the assessments after viewing the visualization helps evaluate the effectiveness of the design relative to communicative intent. Through the assessments, we evaluated both readability accuracy (answering the test question while looking at the visualization) and memorability accuracy (answering the test question after the visualization has been taken away). \revis{Our results demonstrate that while they are useful as a specification, assessments may be even more valuable in the design process as they help identify less effective choices. Because assessments are derived directly from the objectives and insights they can better ensure that the design choices made are good ones. Additionally, running multiple assessments may allow the designer to understand the trade-offs between different goals (i.e., when one visualization cannot `maximize' effectiveness across multiple intents).}

\revis{Finally, it is worth noting that in many situations designers are working in collaborative settings (e.g., with editors, journalists, clients, publishers, etc.). While an expert designer may create or select good visualizations on an independent project, a formal specification `language' may enable better collaborations.}

\subsection{Limitations and Future Directions}
Our study also demonstrates that creating learning objectives for visualization design is a possible way to formulate communicative intent. Although the participants in our study did not create the learning objectives themselves, they could meaningfully use them to choose better designs. As we demonstrate, it is feasible to create these learning objectives.  However, while feasible, generating specifications is not easy. During the process of creating these learning objectives we ourselves needed multiple iterations to create the learning objectives. 
Previous work in this area has also noted that using learning objectives is difficult and needs to be learned and practiced \cite{adar_communicative_2020}. We suggest that more resources be developed to help designers---or those working with them---to create learning objectives for visualizations (just as there are resources in the original domain to help educators create learning objectives in the context of classrooms). \revis{It is also worth acknowledging that many individuals work as visualization designers without any specific training (e.g., a scientist writing an article with figures). A formal specification may help guide these many `novice designers' in building better visualizations.} Even though there is a learning curve to learning objectives, we have shown an advantage to this specification form over the other types that we have studied.

\revis{Communicative visualizations are often embedded in a broader context: articles, presentations, reports. Our goal, in this first study, was to understand the effectiveness of visualizations independently of this additional context. While our specifications were extracted from the original materials, these intents were emphasized differently in the text in ways that could not be easily controlled for. Future studies can clarify the interaction between specification, context, and visualization(s).}

In this study, we recruited crowd workers to rate different visualizations. The reality is that `designers' in our broad sense, are common. Many people must create or pick visualizations without significant training. Thus, helping them specify their intent may lead to better outcomes. However, among this population, we observed a relatively low qualification pass rate. In some sense this is encouraging as it ensured a high level of visual literacy among our participants.   However, we hope to validate our work with other populations. For example, we would like to know if those with more training or experience in design similarly benefit from specifications (and which ones).

To allow for broad participation, we utilized pre-designed graphs instead of focusing on the visualization creation process. In the future, we hope to examine how different types of specifications will affect the design process. We hope to conduct in-depth qualitative design sessions to reveal more about how designers think about communicative intent and how they use it to make design choices. 
Finally, we hope to study further how the combination of learning objectives and assessments can not only guide communicative intent, but can be utilized as part of the design process to inform better design choices.

%% file: sections/07_conclusion.tex
\section{Conclusions}

Communicative intent is an essential part of the visualization design process. Without clear intent, choosing what data to show and how to show it, may lead to worse choices. There are multiple ways to construct a statement of communicative intent (learning objective, insights, assessments). In this work, we identified which specification format was most likely to change baseline preferences for the better. While we found that learning objectives may be the best specification format, we saw improvements over the baseline for the other specification forms as well. Moreover, establishing a communicative intent makes it possible to evaluate the graphs based on the goal. If designers do not have a clear idea of what the visualization should be achieving, then it is hard to meaningfully evaluate the graphs other than just by aesthetics and generic designer guidelines. We encourage designers to articulate their communicative intent to guide more effective design choices. 

%% file: sections/08_ack.tex
\acknowledgments{
We thank our anonymous reviewers for their feedback. We are grateful to the NSF for their support of this work through NSF IIS-1815760.
}

%% file: main.bbl
\begin{thebibliography}{10}

\bibitem{adar_communicative_2020}
E.~Adar and E.~Lee.
\newblock Communicative visualizations as a learning problem.
\newblock {\em IEEE Transactions on Visualization and Computer Graphics},
  27(2):946--956, 2020.

\bibitem{amar_low-level_2005}
R.~Amar, J.~Eagan, and J.~Stasko.
\newblock Low-level components of analytic activity in information
  visualization.
\newblock In {\em {IEEE} Symposium on Information Visualization, 2005.}, pp.
  111--117. {IEEE}, 2005. doi: {{%
10\hspace{.1pt}\discretionary{.}{%
}{.}\hspace{.4pt}1109\discretionary{/}{%
}{/}INFVIS\hspace{.1pt}\discretionary{.}{%
}{.}\hspace{.4pt}2005\hspace{.1pt}\discretionary{.}{%
}{.}\hspace{.4pt}1532136}}


\bibitem{anderson_taxonomy_2004}
L.~W. Anderson and D.~R. Krathwohl, eds.
\newblock {\em {A Taxonomy for Learning, Teaching, and Assessing: A Revision of
  Bloom's Taxonomy of Educational Objectives}}.
\newblock Allyn \& Bacon, New York, 2 ed., December 2001.

\bibitem{bertin1983semiology}
J.~Bertin.
\newblock {\em Semiology of Graphics: Diagrams, Networks, Maps}.
\newblock University of Wisconsin Press, Madison, WI, 1983.

\bibitem{biggs2014evaluating}
J.~B. Biggs and K.~F. Collis.
\newblock {\em Evaluating the quality of learning: The SOLO taxonomy (Structure
  of the Observed Learning Outcome)}.
\newblock Academic Press, 1982.

\bibitem{borkin_beyond_2016}
M.~A. Borkin, Z.~Bylinskii, N.~W. Kim, C.~M. Bainbridge, C.~S. Yeh, D.~Borkin,
  H.~Pfister, and A.~Oliva.
\newblock Beyond memorability: Visualization recognition and recall.
\newblock {\em {IEEE} Transactions on Visualization and Computer Graphics},
  22(1):519--528, 2016. doi: {{%
10\hspace{.1pt}\discretionary{.}{%
}{.}\hspace{.4pt}1109\discretionary{/}{%
}{/}TVCG\hspace{.1pt}\discretionary{.}{%
}{.}\hspace{.4pt}2015\hspace{.1pt}\discretionary{.}{%
}{.}\hspace{.4pt}2467732}}


\bibitem{cairo2012functional}
A.~Cairo.
\newblock {\em The Functional Art: An introduction to information graphics and
  visualization}.
\newblock New Riders, 2012.

\bibitem{card1999readings}
S.~K. Card, J.~D. Mackinlay, and B.~Shneiderman, eds.
\newblock {\em Readings in Information Visualization: Using Vision to Think}.
\newblock Morgan Kaufmann Publishers Inc., San Francisco, CA, USA, 1999.

\bibitem{chang_defining_2009}
R.~Chang, C.~Ziemkiewicz, T.~M. Green, and W.~Ribarsky.
\newblock Defining insight for visual analytics.
\newblock {\em {IEEE} Computer Graphics and Applications}, 29(2):14--17, 2009.
  doi: {{%
10\hspace{.1pt}\discretionary{.}{%
}{.}\hspace{.4pt}1109\discretionary{/}{%
}{/}MCG\hspace{.1pt}\discretionary{.}{%
}{.}\hspace{.4pt}2009\hspace{.1pt}\discretionary{.}{%
}{.}\hspace{.4pt}22}}


\bibitem{chang2010learning}
R.~Chang, C.~Ziemkiewicz, R.~Pyzh, J.~Kielman, and W.~Ribarsky.
\newblock Learning-based evaluation of visual analytic systems.
\newblock In {\em Proceedings of the 3rd BELIV'10 Workshop: BEyond time and
  errors: novel evaLuation methods for Information Visualization}, pp. 29--34,
  2010.

\bibitem{chen2009toward}
Y.~Chen, J.~Yang, and W.~Ribarsky.
\newblock Toward effective insight management in visual analytics systems.
\newblock In {\em 2009 IEEE Pacific Visualization Symposium}, pp. 49--56. IEEE,
  2009.

\bibitem{clark2010graphics}
R.~C. Clark and C.~Lyons.
\newblock {\em Graphics for learning: Proven guidelines for planning,
  designing, and evaluating visuals in training materials}.
\newblock John Wiley \& Sons, 2010.

\bibitem{Diakopoulos20011playable}
N.~Diakopoulos, F.~Kivran-Swaine, and M.~Naaman.
\newblock Playable data: Characterizing the design space of game-y
  infographics.
\newblock In {\em Proceedings of the SIGCHI Conference on Human Factors in
  Computing Systems}, p. 1717–1726. ACM, 2011. doi: {{%
10\hspace{.1pt}\discretionary{.}{%
}{.}\hspace{.4pt}1145\discretionary{/}{%
}{/}1978942\hspace{.1pt}\discretionary{.}{%
}{.}\hspace{.4pt}1979193}}


\bibitem{phillips}
{Economist}.
\newblock The {P}hillips curve may be broken for good, 2017.
\newblock
  https://www.economist.com/graphic-detail/2017/11/01/the-phillips-curve-may-be-broken-for-good.

\bibitem{italy}
{Economist}.
\newblock Italy spends lots fixing old roads, not enough building new ones,
  2018.
\newblock
  https://www.economist.com/graphic-detail/2018/08/15/italy-spends-lots-fixing-old-roads-not-enough-building-new-ones.

\bibitem{covid}
{Economist}.
\newblock American retailers have laid off or furloughed one-fifth of their
  workers, 2020.
\newblock
  https://www.economist.com/graphic-detail/2020/06/03/american-retailers-have-laid-off-or-furloughed-one-fifth-of-their-workers.

\bibitem{few2004show}
S.~Few.
\newblock {\em Show Me the Numbers: Designing Tables and Graphs to Enlighten}.
\newblock Analytics Press, 2004.

\bibitem{forsell_heuristic_2010}
C.~Forsell and J.~Johansson.
\newblock An heuristic set for evaluation in information visualization.
\newblock In {\em Proceedings of the International Conference on Advanced
  Visual Interfaces}, pp. 199--206. {ACM}, 2010. doi: {{%
10\hspace{.1pt}\discretionary{.}{%
}{.}\hspace{.4pt}1145\discretionary{/}{%
}{/}1842993\hspace{.1pt}\discretionary{.}{%
}{.}\hspace{.4pt}1843029}}


\bibitem{gronlund2008gronlund}
N.~E. Gronlund and S.~M. Brookhart.
\newblock {\em Gronlund's writing instructional objectives, 8th ed.}
\newblock Pearson, 2008.

\bibitem{haladyna2012developing}
T.~M. Haladyna.
\newblock {\em Developing and validating multiple-choice test items, 3rd ed.}
\newblock Routledge, 2015.

\bibitem{hearst_evaluating_2016}
M.~A. Hearst, P.~Laskowski, and L.~Silva.
\newblock Evaluating information visualization via the interplay of heuristic
  evaluation and question-based scoring.
\newblock In {\em Proceedings of the 2016 {CHI} Conference on Human Factors in
  Computing Systems}, pp. 5028--5033. {ACM}, 2016. doi: {{%
10\hspace{.1pt}\discretionary{.}{%
}{.}\hspace{.4pt}1145\discretionary{/}{%
}{/}2858036\hspace{.1pt}\discretionary{.}{%
}{.}\hspace{.4pt}2858280}}


\bibitem{hogan2016elicitation}
T.~{Hogan}, U.~{Hinrichs}, and E.~{Hornecker}.
\newblock The elicitation interview technique: Capturing people's experiences
  of data representations.
\newblock {\em IEEE Transactions on Visualization and Computer Graphics},
  22(12):2579--2593, 2016. doi: {{%
10\hspace{.1pt}\discretionary{.}{%
}{.}\hspace{.4pt}1109\discretionary{/}{%
}{/}TVCG\hspace{.1pt}\discretionary{.}{%
}{.}\hspace{.4pt}2015\hspace{.1pt}\discretionary{.}{%
}{.}\hspace{.4pt}2511718}}


\bibitem{hohman2020communicating}
F.~Hohman, M.~Conlen, J.~Heer, and D.~H.~P. Chau.
\newblock Communicating with interactive articles.
\newblock {\em Distill}, 2020.
\newblock https://distill.pub/2020/communicating-with-interactive-articles.
  doi: {{%
10\hspace{.1pt}\discretionary{.}{%
}{.}\hspace{.4pt}23915\discretionary{/}{%
}{/}distill\hspace{.1pt}\discretionary{.}{%
}{.}\hspace{.4pt}00028}}


\bibitem{hollingsed_usability_2007}
T.~Hollingsed and D.~G. Novick.
\newblock Usability inspection methods after 15 years of research and practice.
\newblock In {\em Proceedings of the 25th annual {ACM} international conference
  on Design of communication - {SIGDOC} '07}, p. 249. {ACM}, 2007. doi: {{%
10\hspace{.1pt}\discretionary{.}{%
}{.}\hspace{.4pt}1145\discretionary{/}{%
}{/}1297144\hspace{.1pt}\discretionary{.}{%
}{.}\hspace{.4pt}1297200}}


\bibitem{hullman_benefitting_2011}
J.~Hullman, E.~Adar, and P.~Shah.
\newblock Benefitting {InfoVis} with visual difficulties.
\newblock {\em {IEEE} Transactions on Visualization and Computer Graphics},
  17(12):2213--2222, 2011. doi: {{%
10\hspace{.1pt}\discretionary{.}{%
}{.}\hspace{.4pt}1109\discretionary{/}{%
}{/}TVCG\hspace{.1pt}\discretionary{.}{%
}{.}\hspace{.4pt}2011\hspace{.1pt}\discretionary{.}{%
}{.}\hspace{.4pt}175}}


\bibitem{hullman_visualization_2011}
J.~Hullman and N.~Diakopoulos.
\newblock Visualization rhetoric: Framing effects in narrative visualization.
\newblock {\em {IEEE} Transactions on Visualization and Computer Graphics},
  17(12):2231--2240, 2011. doi: {{%
10\hspace{.1pt}\discretionary{.}{%
}{.}\hspace{.4pt}1109\discretionary{/}{%
}{/}TVCG\hspace{.1pt}\discretionary{.}{%
}{.}\hspace{.4pt}2011\hspace{.1pt}\discretionary{.}{%
}{.}\hspace{.4pt}255}}


\bibitem{isenberg_grounded_2008}
P.~Isenberg, T.~Zuk, C.~Collins, and S.~Carpendale.
\newblock Grounded evaluation of information visualizations.
\newblock In {\em Proceedings of the 2008 conference on {BEyond} time and
  errors novel {evaLuation} methods for Information Visualization - {BELIV}
  '08}, pp. 1--8. {ACM} Press, 2008. doi: {{%
10\hspace{.1pt}\discretionary{.}{%
}{.}\hspace{.4pt}1145\discretionary{/}{%
}{/}1377966\hspace{.1pt}\discretionary{.}{%
}{.}\hspace{.4pt}1377974}}


\bibitem{kim2017explaining}
Y.-S. Kim, K.~Reinecke, and J.~Hullman.
\newblock Explaining the gap: Visualizing one's predictions improves recall and
  comprehension of data.
\newblock In {\em Proceedings of the 2017 CHI Conference on Human Factors in
  Computing Systems}, pp. 1375--1386, 2017.

\bibitem{KindlmannAlgebraicVisDesign2014}
G.~Kindlmann and C.~Scheidegger.
\newblock An algebraic process for visualization design.
\newblock {\em IEEE Transactions on Visualization and Computer Graphics
  (Proceedings VIS 2014)}, 20(12):2181--2190, 2014. doi: {{%
10\hspace{.1pt}\discretionary{.}{%
}{.}\hspace{.4pt}1109\discretionary{/}{%
}{/}TVCG\hspace{.1pt}\discretionary{.}{%
}{.}\hspace{.4pt}2014\hspace{.1pt}\discretionary{.}{%
}{.}\hspace{.4pt}2346325}}


\bibitem{mackinlay_automating_1986}
J.~Mackinlay.
\newblock Automating the design of graphical presentations of relational
  information.
\newblock {\em {ACM} Transactions on Graphics ({TOG})}, 5(2):110--141, 1986.
  doi: {{%
10\hspace{.1pt}\discretionary{.}{%
}{.}\hspace{.4pt}1145\discretionary{/}{%
}{/}22949\hspace{.1pt}\discretionary{.}{%
}{.}\hspace{.4pt}22950}}


\bibitem{north_toward_2006}
C.~North.
\newblock Toward measuring visualization insight.
\newblock {\em {IEEE} Computer Graphics and Applications}, 26(3):6--9, 2006.
  doi: {{%
10\hspace{.1pt}\discretionary{.}{%
}{.}\hspace{.4pt}1109\discretionary{/}{%
}{/}MCG\hspace{.1pt}\discretionary{.}{%
}{.}\hspace{.4pt}2006\hspace{.1pt}\discretionary{.}{%
}{.}\hspace{.4pt}70}}


\bibitem{nowak2018micro}
S.~Nowak, L.~Bartram, and T.~Schiphorst.
\newblock A micro-phenomenological lens for evaluating narrative visualization.
\newblock In {\em 2018 IEEE Evaluation and Beyond-Methodological Approaches for
  Visualization (BELIV)}, pp. 11--18. IEEE, 2018.

\bibitem{popham2003test}
W.~J. Popham.
\newblock {\em Test better, teach better: The instructional role of
  assessment}.
\newblock ssociation forSupervision and Curriculum Development, 2003.

\bibitem{raible2016writing}
J.~Raible, L.~Bennett, and K.~Bastedo.
\newblock Writing measurable learning objectives to aid successful online
  course development.
\newblock {\em Int. J. for Scholarship of Technology Enhanced Learning}, 1(1),
  2016.

\bibitem{saket2018task}
B.~Saket, A.~Endert, and {\c{C}}.~Demiralp.
\newblock Task-based effectiveness of basic visualizations.
\newblock {\em IEEE transactions on visualization and computer graphics},
  25(7):2505--2512, 2018.

\bibitem{tufte2001visual}
E.~Tufte.
\newblock {\em The Visual Display of Quantitative Information}.
\newblock Graphics Press, 2001.

\bibitem{tufte2006beautiful}
E.~R. Tufte.
\newblock {\em Beautiful evidence}.
\newblock Graphics Press Cheshire, CT, 2006.

\bibitem{ware2019information}
C.~Ware.
\newblock {\em Information visualization: perception for design}.
\newblock Morgan Kaufmann, 2019.

\bibitem{wiggins2005understanding}
G.~Wiggins and J.~McTighe.
\newblock {\em Understanding by Design}.
\newblock Association for Supervision and Curriculum Development, 2005.

\bibitem{wong2013wall}
D.~Wong.
\newblock {\em The Wall Street Journal Guide to Information Graphics: The Dos
  and Dont's of Presenting Data, Facts, and Figures}.
\newblock W. W. Norton, Incorporated, 2013.

\bibitem{yi2008}
J.~S. Yi, Y.-a. Kang, J.~T. Stasko, and J.~A. Jacko.
\newblock Understanding and characterizing insights: How do people gain
  insights using information visualization?
\newblock In {\em Proceedings of the 2008 Workshop on BEyond Time and Errors:
  Novel EvaLuation Methods for Information Visualization}, BELIV '08. ACM,
  2008. doi: {{%
10\hspace{.1pt}\discretionary{.}{%
}{.}\hspace{.4pt}1145\discretionary{/}{%
}{/}1377966\hspace{.1pt}\discretionary{.}{%
}{.}\hspace{.4pt}1377971}}


\end{thebibliography}
